\RequirePackage{rotating}
\documentclass[twocolumn,times,10pt]{aastex61}
\usepackage[USenglish,american]{babel}
\usepackage{graphicx}
\usepackage{multirow}
\usepackage{amsmath}

\usepackage{soul}

\accepted{March 19, 2018} 
\submitjournal{ApJ}

\shorttitle{A nearby hypervelocity supernova remnant}
\shortauthors{Raddi et al.}

\begin{document}

\title{Further insight on the hypervelocity white dwarf, LP\,40$-$365 (GD\,492):\\
a nearby emissary from a single-degenerate Type Ia supernova}

\correspondingauthor{R. Raddi}
\email{roberto.raddi@fau.de}

\author[0000-0002-9090-9191]{R. Raddi}
\affiliation{University of Warwick, Department of Physics, Gibbet Hill Road, Coventry, CV4 7AL, United Kingdom}

\author{M. A. Hollands}
\affiliation{University of Warwick, Department of Physics, Gibbet Hill Road, Coventry, CV4 7AL, United Kingdom}

\author{D. Koester}
\affiliation{Universit\"at Kiel, Institut f\"ur Theoretische Physik und Astrophysik, 24098, Kiel, Germany}

\author[0000-0002-2761-3005]{B. T. G\"ansicke}
\affiliation{University of Warwick, Department of Physics, Gibbet Hill Road, Coventry, CV4 7AL, United Kingdom}

\author{N. P. Gentile Fusillo}
\affiliation{University of Warwick, Department of Physics, Gibbet Hill Road, Coventry, CV4 7AL, United Kingdom}

\author[0000-0001-5941-2286]{J. J. Hermes}
\altaffiliation{Hubble Fellow}
\affiliation{University of North Carolina, Department of Physics and Astronomy, Chapel Hill, NC - 27599-3255, USA}

\author[0000-0002-9538-5948]{D. M. Townsley}
\affiliation{University of Alabama, Department of Physics and Astronomy,  Tuscaloosa, AL, USA}

\begin{abstract}\noindent The recently discovered hypervelocity white dwarf LP\,40$-$365 (aka GD\,492) has been suggested as the outcome of the failed disruption of a white dwarf in a sub-luminous Type~Ia supernova (SN~Ia). We present new observations confirming GD\,492 as a single star with unique spectral features. Our spectroscopic analysis suggests that a helium-dominated atmosphere, with $\simeq 33$\% neon and 2\% oxygen by mass, can reproduce most of the observed properties of this highly unusual star. Although our atmospheric model contrasts with the previous analysis in terms of dominant atmospheric species, we confirm that the atmosphere of GD\,492 is strongly hydrogen deficient, $\log{(\mathrm{H/He})}<-5$, and displays traces of eleven other $\alpha$ and iron-group elements (with sulfur, chromium, manganese, and titanium as new detections), indicating nuclear processing of carbon and silicon. We measure a manganese-to-iron ratio seven times larger than Solar. While the observed abundances of GD\,492 do not fully match any predicted nuclear yields of a partially-burned supernova remnant, the manganese excess strongly favors a link with a single-degenerate SN~Ia event over alternative scenarios. 
\end{abstract}


\section{Introduction} \label{sec:intro}
Type~Ia supernovae (SNe~Ia) are luminous transient events, which are interpreted as the explosions of carbon/oxygen (C/O) core white dwarfs that have accreted enough mass from a companion star to trigger runaway nuclear reactions \citep{nomoto97}. SN~Ia light-curves are regarded as key standardizable candles for constraining the properties of the Universe \citep{riess98,perlmutter99}.

Recent large-area transient searches have revealed a complex variety of light curve shapes and spectral features, which are explained with underlying differences in the explosion mechanisms and architecture of the binary progenitors, e.g. Type Iax \citep[][]{foley13}, Type .Ia \citep[][]{shen10}, or calcium-rich transients \citep[][]{kaliswal12}.  Observation-based evidence and theoretical arguments support multiple progenitor scenarios, broadly defined as single and double-degenerate systems, i.e.\ containing one or two white dwarfs, respectively \citep{wang12,maoz14}. The relative importance of these two channels remains intensively discussed in the context of observationally determined supernova rates and delay time distributions \citep{mannucci06,maoz10}, and it is interlinked to the possible explosion mechanisms that need to match observable quantities such as light-curves, nuclear yields, and total energy output \citep[see e.g.\ 3D hydrodynamic simulations of pure deflagrations, deflagration-to-detonation transitions, delayed detonations, and double detonations;][]{ropke07,kromer10,seitenzahl13b,fink14}.

Recently, \citet{vennes17} reported the discovery of a nearby, chemically-peculiar white dwarf, LP\,40$-$365, that is escaping the Milky Way with a radial velocity of $\simeq 500$\,km\,s$^{-1}$. This star had long been known as a relatively bright ($V=15.5$\,mag) star with a moderately high proper motion \citep[$\approx 160$~mas/yr;][]{luyten70}. It was later proposed as a white dwarf suspect \citep[GD\,492\footnote{Although the \citet{giclas70} designation has already been used in the white dwarf literature, we will use the \citet{luyten70} name for continuity with \citet{vennes17}.};][]{giclas70} and more recently confirmed as a likely candidate white dwarf \citep[][]{limoges15}; however,  there was no spectroscopic classification prior to \citet{vennes17}. Based on low- and high-resolution spectroscopy, the authors argued for this star to be the surviving remnant of a partially-burned (hereafter, ``unburnt'') white dwarf from a subluminous SN~Ia, i.e.\ a compact object that did not entirely disrupt in an explosion with a transient light curve of the likes of SNe~Iax \citep[][these authors also call the remnants ``bound'']{jordan12,kromer13}.

We report new observations of LP\,40$-$365, which were obtained as part of a spectroscopic follow-up program targeting nearby white dwarfs \citep[][]{raddi17}, and present a spectroscopic analysis of this highly unusual star based on our independent stellar atmosphere code. 

\section{observations} \label{sec:two}
\subsection{Spectroscopy} \label{sec:two-1}

\begin{figure}
\centering
\includegraphics[width=\linewidth]{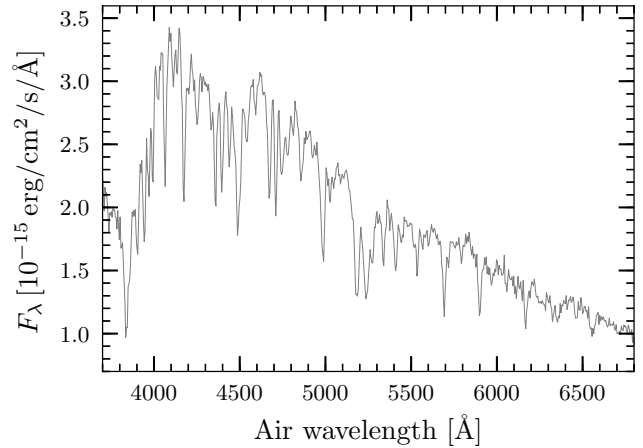}
\caption{Low-resolution Asiago identification spectrum. 
\label{f:fig01}}
\end{figure}

We acquired three 20-min spectra of LP\,40$-$365 on 2016 December 26, with the 1.82-m Copernico Telescope at the Asiago Observatory in Italy. Using the Asiago Faint Object Spectrograph and Camera (AFOSC) equipped with the VPH\,\#7 grism \citep{zanutta14} and a 1.25-arcsec slit. The average of these low-resolution (13~\AA) spectra has a signal-to-noise ratio (S/N) of $\approx 50$ in the 3500--7300~\AA\ range. The unusual spectral appearance of LP\,40$-$365 was already evident from these data (Fig\,\ref{f:fig01}). 

On 2017 March 1 and April 2, we secured higher resolution spectroscopy (2\,\AA) with the Intermediate dispersion Spectrograph and Imaging System (ISIS) mounted on the 4.2-m William Hershel Telescope (WHT) in La Palma (Spain). We used the D5300 beam-splitting dichroic and the R600B/R600R grisms with a 1-arcsec slit. In both runs, we took 10--15\,min exposures, modifying the central wavelengths in the blue and red setups, in order to cover the 3200--5300~\AA\ and 5700--9100~\AA\ wavelength ranges. Exposing for a total of $\approx 1.4$~hr, the co-added spectra have a S/N$> 30$ per pixel.  We re-observed LP\,40$-$365 in service mode at the WHT on 2017 July 6 and August 10, obtaining $21\times10$-min exposures using the R1200B/R1200R grisms and a 1-arcsec slit, to cover the 4500--5400 and 8100--9000~\AA\ wavelength ranges at 1\,\AA\ resolution. Also with this higher-resolution setup, we obtained a combined spectrum with S/N$ > 30$ from the 1.5~hr July run. Poorer quality data were obtained during the 2~hr August run, when LP\,40$-$365 was observed at airmass $> 3$. The journal of the observations, including the instrument setups and the spectral-range coverage, is listed in Table\,\ref{t:t01}

In all the WHT runs we took particular care to choose suitable standard stars to achieve an acceptable relative flux calibration and to remove the effect of atmospheric telluric bands at red wavelengths. We reduced the raw frames, optimally-extracted the one-dimensional spectra, and applied a wavelength and flux calibration by using the {\sc starlink}\footnote{\url{http://starlink.eao.hawaii.edu/starlink}} suite of software that includes {\sc pamela} and {\sc molly}\footnote{\url{http://www.warwick.ac.uk/go/trmarsh/software/}} \citep{marsh89}. The co-added WHT/ISIS spectra are analyzed in Section\,\ref{sec:three}.

\subsection{Photometry} \label{sec:two-2}
\begin{table*}
\centering
\caption{Journal of the spectroscopic observations.\label{t:t01}}
{\footnotesize 
\begin{tabular}{lcccccccc}
\hline
\hline
Telescope & Instrument & Grating & Night & N. exp & Tot. Exp. & Resolution & Range \\
      &           &            & & & [sec] & [\AA] & [\AA]\\
\hline
  Copernico & AFOSC & VPH \#7  & 2016-12-28  & 3 &  3600 & 13 & 3600--7300 \\
\hline
\multirow{8}{*}{WHT} & \multirow{8}{*}{ISIS}  & \multirow{4}{*}{600B/600R} & \multirow{2}{*}{2017-03-01}  & \multirow{2}{*}{3} &  \multirow{2}{*}{2700} & \multirow{4}{*}{2}  & 3200--5300 \\
  &   &   &  &  &  & & 5700--9100 \\
 &  &   & \multirow{2}{*}{2017-04-02}  & \multirow{2}{*}{4} &  \multirow{2}{*}{2400} & & 3200--5300 \\
  &   &   &  & & & & 5700--9100 \\
\cline{3-8}
 &  & \multirow{4}{*}{1200B/1200R}  & \multirow{2}{*}{2017-07-06}& \multirow{2}{*}{9} &  \multirow{2}{*}{5400}  & \multirow{4}{*}{1} & 4500--5400\\
  &   &   &  & & & &  8100--9000\\
 &  &    & \multirow{2}{*}{2017-08-10}&  \multirow{2}{*}{12} &  \multirow{2}{*}{7200}  &  & 4500--5400\\
  &   &   &  & & & &  8100--9000\\
\hline
\end{tabular}}
\end{table*}

We acquired four hours of fast photometry on 2017 May 22 with the fully autonomous 2-m Liverpool Telescope, using the fast-readout camera RISE requested with a reactive-time mode proposal.  We took  5-sec exposures  with a $2\times2$ binning, achieving S/N$ = 30$ per image. We performed differential aperture photometry with respect to USNO-B1.0 1643--0083234 ($B=13.8$, $R=12.9$) using the SExtractor \citep{bertin+arnouts96-1} based pipeline described by \citet{gaensickeetal04-1}. We did not detect any noticeable feature in the light curve of LP\,40$-$365 apart from a long-term trend caused by differential extinction in the very broad non-standard $V+R$-band filter installed in the RISE camera. A discrete Fourier transform computed from the data did not reveal any significant periodic signal with an amplitude limit of $0.002$\,mag. We therefore conclude that the LT photometry does not provide any evidence of either binarity (eclipses, ellipsoidal modulation or reflection effect) or short-period variability.

\begin{figure*}
\begin{sideways}
\begin{minipage}[c]{21cm}
\centering
\includegraphics[width=\textwidth]{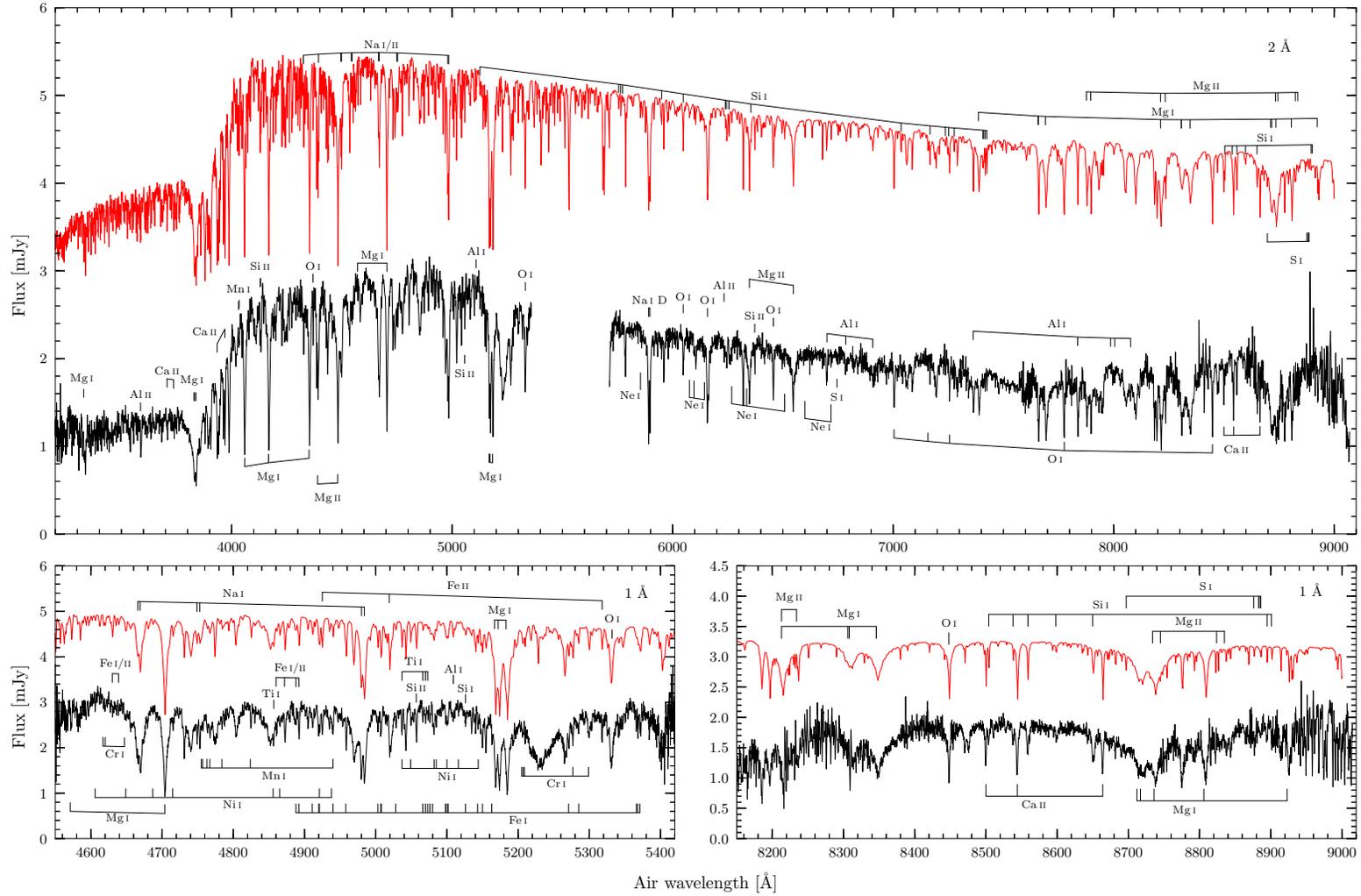}
{\caption{WHT/ISIS spectra of LP\,40$-$365 (black) with our best-fitting synthetic spectrum (red), vertically-offset by 1.5--2.5\,mJy in the bottom and upper panels, respectively. The data have been blue-shifted into the rest frame by $500$\,km\,s$^{-1}$. We note that some misalignments in relative flux levels between observed and modeled spectra may result from poorer calibration at the CCD edges. We display the averaged 2-\AA\ WHT/ISIS spectra in the top panel, and the blue and red averaged 1-\AA\ spectra in the bottom panels. Relevant transitions are labeled in each panel. We note the presence of the $\simeq40$-\AA\ wide feature at 5250\,\AA\ in the observed spectra, which is one of the strongest features that we are not able to reproduce with our model spectrum (see text in Section\,\ref{sec:three} for details). \label{f:fig02}}}
\end{minipage}
\end{sideways}
\end{figure*}
\section{Spectroscopic analysis}\label{sec:three}

The low-resolution Asiago/AFOSC spectrum (Fig.\,\ref{f:fig01}) and the broad band photometry (Table\,\ref{t:t02}) are already sufficient to confirm the main properties of this object reported by \citet{vennes17}, i.e. $T_{\rm eff} \sim 10\,000$~K and $\log{g} \sim 5.5$. Even at this very low resolution, the Mg 3830\,\AA\ and 5170\,\AA\ lines are recognizable and indicate a large radial velocity of $\sim 500$\,km\,s$^{-1}$, also in agreement with the findings of \citep{vennes17}. 

Inspecting the combined 1-\AA\ resolution spectrum from July (which has a better quality than our August data), and the 2-\AA\ resolution WHT/ISIS spectrum (Fig\,\ref{f:fig02}), which almost contiguously spans  3200--9100\,\AA, we detect absorption features of the elements that \citet{vennes17} noticed in their data (which was limited to a smaller wavelength range 3850--6850\,\AA), i.e. O, Ne, Na, Mg, Al, Si, Ca, Fe, and Ni. In addition, we identify in our spectra transitions of S, Cr, Ti, and Mn. The detection of Mn is particularly important for the discussion of the possible nature of LP\,40$-$365 (see Sect.\,\ref{sec:four})

We also note the steep flux decrease between 3780 and 4000\,\AA\ that, at a first glance, resembled the Balmer jump, but it is instead due to the photoionization absorption of Mg, clearly visible as a result of the extreme abundance of this element. 

Below, we present our atmospheric analysis of the WHT/ISIS spectroscopy, carried out with our model atmosphere code described in \citet{koester10}, updated to the most recent physics. We first determine the fundamental atmospheric parameters ($T_\mathrm{eff}$, $\log g$, and the dominant atmospheric species) and then derive the abundances of the trace elements.

\subsection{Effective temperature, surface gravity, and dominant atmospheric species\label{sec:atmopar}}

\citet{vennes17} found that a O/Ne-dominated atmosphere ($\approx 40$--$50/30$--$50$\% by number, respectively, with $<15$\% He by number) best reproduces their observations, based on the absence of H and He lines. Having a larger wavelength coverage, although with lower resolution, we tested the effect of different dominant atmospheric elements on the overall spectral appearance.  We produced a grid of synthetic spectra covering $T_{\rm eff} = 8300$--$10\,100$\,K, $\log{g} = 5$--$6.5$, using in turn He, O, Ne, and Mg, as dominant species. We excluded H and C as the possible dominant atmospheric components, given that Balmer lines and ${\rm C}_2$ Swan bands are not detected. The underlying stellar structure takes into account the methods of \citet{koester10} with updated  physics. Cool white dwarfs such as LP\,40$-$365 have convective atmospheres, which we model with a mixing length parameter, ${\rm ML}2/\alpha = 1.25$. The model atmosphere is convective below $\tau_{\rm Ross} = 1$, but we note that the atmospheric structure and the emerging flux above this value are only marginally different between models with and without convection (like that adopted by \citealt{vennes17}). The full extent of the convection zone is unknown, as the present code is optimized for a single dominant-element, containing just traces of other elements, while in this case the metal abundances are large enough that they could have an effect. In general, we note that in cool, low-mass white dwarfs, such as LP\,40$-$365, the convection zone could reach down to the core \citep{fontaine76}, thus favoring the mixing of heavier elements.  We will discuss further implications of this in Section~\ref{sec:four}.

\begin{figure}
\centering
\includegraphics[width=\linewidth]{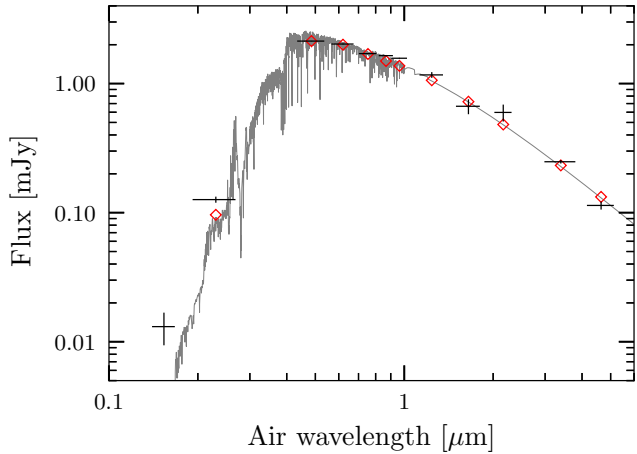}
\caption{Spectral energy distribution of LP\,40$-$365.
The best-fitting model (gray) is compared to broad-band photometry from {\em Galex}, Pan-STARRS, 2MASS, and {\em WISE} (black circles with error-bars). The model spectrum is convolved with a Gaussian filter ($\sigma = 30$\,\AA), normalized to the Pan-STARRS $g$-band \citep{flewelling16},  extended to infrared wavelengths with a $T = 8900$\,K blackbody, and reddened by $E(B-V) = 0.02$ \citep[total line-of-sight reddening, estimated via the dust maps by][]{sfd98}. The diamond symbols represent
the synthetic magnitudes in the filter band passes.\label{f:fig03}}
\end{figure}

Our best-fit model was numerically identified via $\chi^{2}$ minimization, first taking into account $T_{\rm eff}$, $\log{g}$, and the dominant atmospheric element, and then determining the abundances of individual trace elements. To validate the quality of the fit, we examined a range of features,  which included the non-detection of the He\,{\sc i} 5877\,\AA\ line,  the strong Mg\,{\sc i} lines between 4000--4800\,\AA, the strong Mg\,{\sc ii} 4482\,\AA\ and other Mg\,{\sc ii} lines, the flux decrement at the Mg\,{\sc i} photo-ionization edge (3780--4000\,\AA), and the ionization ratio between Mg\,{\sc i}/Mg\,{\sc ii} and Ca\,{\sc i}/Ca\,{\sc ii}. The best-fit model is found at $T_{\rm eff} = 8900 \pm 600$~K and $\log{g} = 5.5 \pm 0.5$. This model has a He-dominated atmosphere, with 10\% of Ne by number, which is in clear contrast with the \citet{vennes17} interpretation.  The errors represent a 2-$\sigma$ uncertainty, and the $T_{\rm eff}$ we measure is at 4-$\sigma$ from the \citet{vennes17} result. We compare the best-fit model to our WHT/ISIS spectra in Fig.\,\ref{f:fig02}, and to the spectral energy distribution (SED) in Fig.\,\ref{f:fig03}, showing a good agreement with the available broad-band photometry from the literature too (Table\,\ref{t:t02}). 

Our best-fit model still includes a weak He\,{\sc i}~5877\,\AA\ line, which reaches 3-$\sigma$ above the noise level of the observed spectrum, in the considered range of $T_{\rm eff}$.  Given the lack of available theories and data for the He-line broadening by neutral atoms in white dwarf atmospheres of such low $T_{\rm eff}$ and $\log{g}$, basic assumptions on the line-profiles and extrapolation from the available computations at room temperature have to be taken into account \citep[for a comprehensive description, see][]{koester10}. With this caveat in mind, we note that uncertainties in the theoretical modeling of He lines may either lead  to systematic overestimates or underestimates of line strengths. 

\begin{table}
\centering
\caption{Broad-band photometry and synthetic fluxes.\label{t:t02}}
\begin{tabular}{lccc}
\hline
\hline
Source & Wavelength & Observed flux  & Model flux \\
& [$\mu$m]&  [mJy] & [mJy] \\
\hline
{\em Galex$^{\rm a}$} FUV  &$0.15$& $0.013$~$\pm~0.004$ & \\
{\em Galex} NUV  &$0.23$& $0.126$~$\pm~0.007$ & 0.096 \\
Pan-STARRS$^{\rm b}$  $g$  &$0.49$& $2.133$~$\pm~0.006$ & 2.133 \\
Pan-STARRS  $r$  &$0.62$& $2.024$~$\pm~0.017$ & 2.000 \\
Pan-STARRS  $i$  &$0.75$& $1.707$~$\pm~0.011$ & 1.694 \\
Pan-STARRS  $y$  &$0.87$& $1.571$~$\pm~0.012$ & 1.372 \\
Pan-STARRS  $z$  &$0.96$& $1.645$~$\pm~0.003$ & 1.495 \\ 
2MASS$^{\rm c}$ $J$	 &$1.2$ & $1.17$~$\pm~0.06$ & 1.06 \\  
2MASS $H$	 &$1.7$ & $0.67$~$\pm~0.09$ & 0.72 \\  
2MASS $Ks$	 &$2.1$ & $0.60$~$\pm~0.09 $  & 0.48 \\  
{\em WISE$^{\rm d}$} $W1$    &$3.4$& $0.248$~$\pm~0.008$ & 0.232  \\
{\em WISE} $W2$   &$4.6$ & $0.114$~$\pm~0.008$ & 0.132 \\
\hline
\multicolumn{3}{l}{$^{\rm a}$\citet{morrissey07}; $^{\rm b}$\citet{flewelling16};}\\
\multicolumn{3}{l}{$^{\rm c}$\citet{skrutskie06}; $^{\rm d}$\citet{wright10}.}\\
\end{tabular}
\end{table}

\begin{figure*}[tb]
\includegraphics[width=\textwidth]{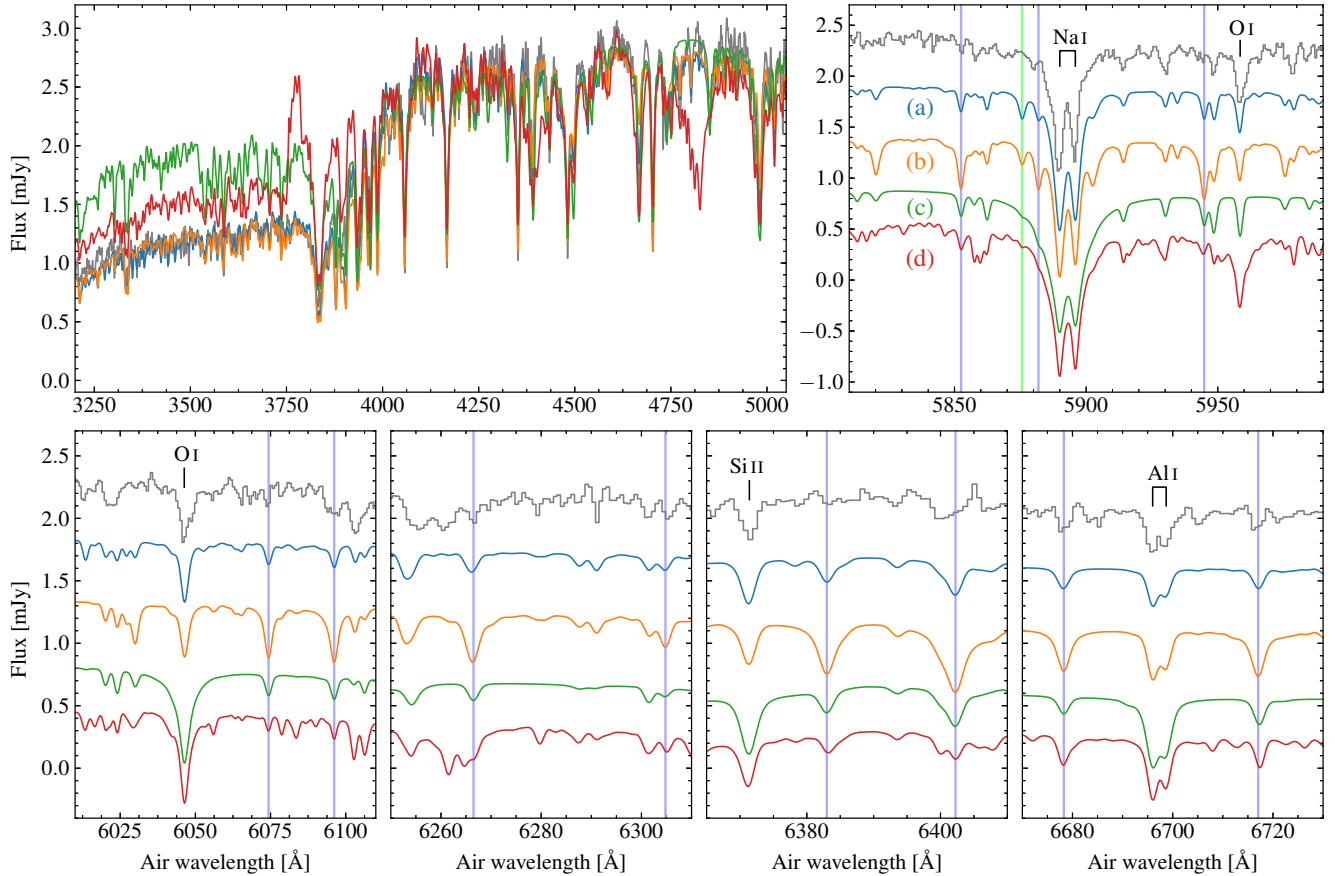}
\caption{Comparison between observed and synthetic spectra in the proximity of the Mg ionization-edge (top left  panel), and He and Ne absorption lines (marked by green and blue vertical lines, respectively, in the other panels). In the top-left  panel, the synthetic spectra are plotted onto the observed spectrum without any offset to highlight the blue-flux excess obtained with higher $T_{\rm eff}$ models. In the other panels, from top to bottom, we display the observed WHT/ISIS spectrum (gray), our best-fit model ((a); blue), a Ne atmosphere with $T_{\rm eff} = 8900$\,K and $\log{g} = 5.5$ ((b); orange) for which we also refitted the element abundances, a Ne atmosphere computed with our code but with element abundances and the atmospheric parameters of \citet{vennes17}, $T_{\rm eff} = 10\,100$\,K and $\log{g} = 5.8$ ((c); green), and the \citet{vennes17} synthetic spectrum ((d); red). The synthetic spectra are convolved to the same instrumental resolution as our data, scaled to the fluxed spectrum between 5000--7000\,\AA\ and vertically offset by steps of $0.5$\,mJy for display purposes. The observed spectra are blue-shifted into the rest frame by 500~km/s$^{-1}$. Strong O\,{\sc i}, Na\,{\sc i} D, Si\,{\sc ii}, and Al\,{\sc i} transitions are labeled above the observed spectrum.\label{f:fig04}}
\end{figure*}
Although we were satisfied with the overall quality of our He-atmosphere fit to the available data,  the results of \citet{vennes17} motivated us to investigate whether a Ne-dominated atmosphere could provide an improved fit. We examined in more detail how synthetic models reproduce key spectral features, such as the non-detection of the He\,{\sc i} 5877\,\AA\ line, the Ne lines, and the Mg\,{\sc ii} series between 3400--5000\,\AA. As outlined above, we could not identify a Ne-dominated model reproducing all the listed features; hence, as an example, we compare our best-fit model with $\log{({\rm Ne}/{\rm He})} = -1$ ((a) in Fig.\,\ref{f:fig04}) against two Ne-dominated models computed with our atmosphere code: the first ((b) in Fig.\,\ref{f:fig04}) has $T_{\rm eff} = 8900$\,K, $\log{g} = 5.5$, $\log{({\rm He}/{\rm Ne})}=-0.5$ \citep[the lower limit on the He abundance from][]{vennes17}, and all the major trace elements adjusted to give a reasonable fit; the second ((c) in Fig.\,\ref{f:fig04}) has $T_{\rm eff} = 10\,100$\,K, $\log{g} = 5.80$, and metal composition as in \citet{vennes17}. Finally, we obtained the synthetic spectrum of \citet{vennes17} ((d) in Fig\,\ref{f:fig04}). It is evident that models (a) and (b), having the same $T_{\rm eff}$ and $\log{g}$ well fit the observed spectrum in the top-left panel of Fig.\,\ref{f:fig04}; however, model (b) over-predicts the Ne line-strengths. In contrast, model (c) well fits the Ne lines, but does not reproduce the intensity of the O, Na, Si, Al (all labeled in Fig.\,\ref{f:fig04}) and the Mg series. Model (c) has the same parameters as the original Vennes et al. model, (d), but we note some discrepancies between the two, which are likely due to the different underlying physics in the spectral synthesis codes. We note that model (d) contains line-blends that are not detected in the data (e.g. the strong absorption band at 4800\,\AA, in the top-left  panel of Fig.\,\ref{f:fig04}). Both model (c) and (d) do not perform well in correspondence of the Mg ionization-edge, over-predicting the flux below 4000\,\AA\ by $\sim 30$\%, implying that our $T_{\rm eff} = 8900$\,K is probably more representative of LP\,40$-$365's atmosphere. The extended coverage at short wavelengths of our data compared to the observations presented by \citet{vennes17} provides an improved constraint on the flux of LP\,40$-$365 at short wavelengths, and therefore on the $T_{\rm eff}$ of the star.  It is worth noting that our fitting procedure and that of \citet{vennes17} differ. We determine our best-fit by comparing the fluxed spectrum to a wide grid of models that cover the parameter space mentioned above, in contrast, \citet{vennes17} adopt an iterative best-fitting procedure that uses continuum-normalized spectra. Both procedures can lead to outweigh some spectral features, i.e. the strength of the He\,{\sc i}\,5877~\AA\ line in our case, or the overall SED in \citet{vennes17}. 

Thus, despite the non-detection of the He\,{\sc i} 5877\,\AA\ line, weighing up all evidence, we argue that the analysis of our data with our model atmosphere code favors a He-dominated atmosphere compared to the \citet{vennes17} O/Ne atmosphere. We note, however, that given the large anti-correlated uncertainties (see next section), the Ne/He ratio of our best-fit model  allows for 20--50\% of Ne by mass, or 5--20\% by number abundance. 

As final remark, we note that our model spectrum does not contain some of the absorption lines or blends that are visible in our low- and intermediate-resolution spectra, or in the \citet{vennes17} data.  The strongest of such features missing in the model is the broad absorption between 5200--5300\,\AA\ that \citet{vennes17} interpreted as Mg\,{\sc i}~$3{\rm p}^{3}$ resonance, though their synthetic spectrum does not contain the observed absorption feature. Given the exotic nature of this star, extremely rich in metals, we stress that models including unexplored physics  will be necessary, e.g.\ taking into account improved photo-ionization cross-sections or the presence of metastable molecular states for the most common elements we use in our analysis (e.g.\ He, O, Ne, Mg), for which the relevant atomic and molecular data are not yet available.

\subsection{The unique atmospheric composition of LP\,40$-$365} \label{sec:three-1}

Inspecting the WHT/ISIS spectra we detect transitions of 13 atomic species that, including Ne, are Mg, O, Na, Si, Ca, Fe, Al, S, Cr, Ti, Mn, and Ni, sorted by average line strength.  This list includes all elements that were identified by \citet{vennes17}, plus the additional detection of S, Cr, Ti, and Mn. 
Starting from the atmospheric parameters as derived above in Sect.\,\ref{sec:atmopar}, i.e. $T_\mathrm{eff}=8900$\,K, $\log g=5.5$, $\log{\mathrm{Ne/He}}=-1$, we proceeded to determine the photospheric abundances of the 13 detected species. Each element was individually assessed by comparing the effect of incremental 0.1-dex changes in atomic abundances on the relative intensity of observed and synthetic spectral lines. Based on the absence of Balmer lines, we could set a very stringent limit on the H abundance, $\log{({\rm H}/{\rm He})} < -5$. We also assessed the upper limits for the remaining most common elements up to Ni. The number abundances and the upper limits for all the elements are given as $\log{({\rm Z}/{\rm Total})}$, in Table\,\ref{t:t03} 

We note that considering individual lines or groups of lines would result in slight different abundances. For example, by just fitting the Ne lines that are visible in the \citet{vennes17} spectra, we obtain a Ne abundance lower by $0.3-0.5$\,dex than reported in Table\,\ref{t:t03}. These variations are likely due to uncertainties in the available atomic data. The results given in Table\,\ref{t:t03} are obtained from the best fit of  all detected lines. Finally, we note that element abundances correlate with $T_{\rm eff}$,  thus we estimated the systematic uncertainties by taking the abundances at the 1-$\sigma$ $T_{\rm eff}$ range, corresponding to an average variation of $\pm0.25$\,dex. The systematic uncertainty is adopted whenever abundance ratios are considered, except for Ne, which anti-correlates by a similar factor. 
\begin{table}
\caption{Logarithmic number abundances of detected elements and upper limits. The errors represent a 2-$\sigma$ uncertainty, i.e. 95\% confidence regions. The uncertainty associated to the dominant species, He, reflects the variation of the other elements (mainly Ne) with respect to the total. \label{t:t03}}
\centering
\begin{tabular}{cDcD}
\hline
\hline
\noalign{\smallskip}
Element  & \multicolumn2c{$\log{({\rm Z}/{\rm Total})}$} & Element  & \multicolumn2c{$\log{({\rm Z}/{\rm Total})}$}\\
\noalign{\smallskip}
\hline
\decimals
 H & $<-5.05$           & Cl & $<-5.85$  \\
He & $-0.05$~$\pm0.05$  & Ar & $<-4.35$  \\   
 C & $<-4.55$           & Ca & $-6.55$~$\pm~0.30$ \\
 O & $-2.15$~$\pm~0.20$ & Sc & $<-8.55$  \\
Ne & $-1.05$~$\pm~0.30$ & Ti & $-7.07$~$\pm~0.15$ \\
Na & $-4.07$~$\pm~0.15$ & V  & $<-6.35$  \\
Mg & $-3.05$~$\pm~0.15$ & Cr & $-6.33$~$\pm~0.20$  \\
Al & $-4.36$~$\pm~0.20$ & Mn & $-6.15$~$\pm~0.20$  \\
Si & $-4.53$~$\pm~0.20$ & Fe & $-4.9$~$\pm~0.30$  \\
P  & $<-4.55$           & Ni & $-5.94$~$\pm~0.20$  \\
S  & $-5.19$~$\pm~0.15$ & \\
\hline
\end{tabular}
\end{table}

\begin{figure}
\centering
\includegraphics[width=\linewidth]{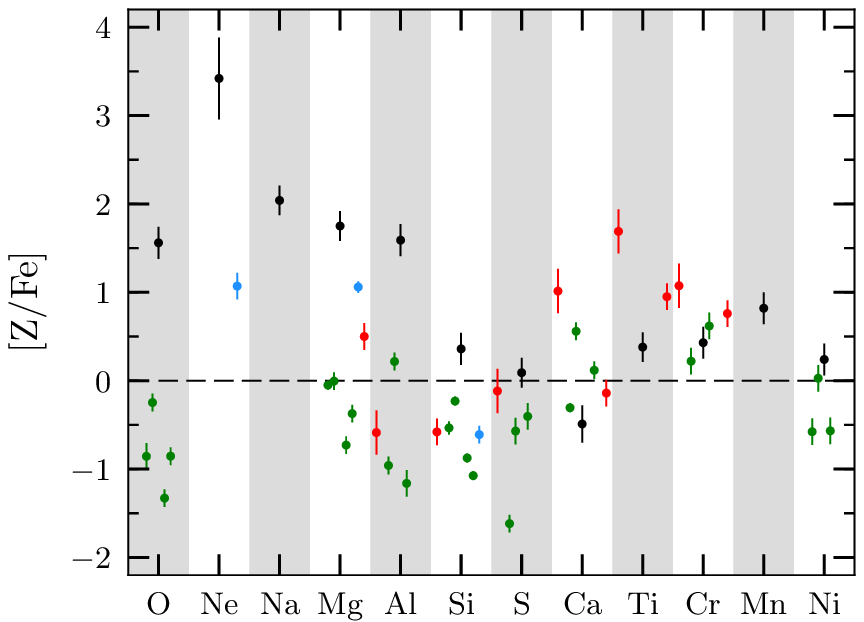}
\caption{The abundances of detected nuclear-burning products, relative to iron(black symbols), are compared to extremely low-mass white dwarfs \citep[red;][]{hermesetal14-1,gianninasetal14-1}, metal-polluted white dwarfs \citep[green;][]{gaensicke12}, and the O-atmosphere white dwarf (light blue) of  \citet{kepleretal16}. The number abundances are plotted relative to Fe and are normalized to Solar values \citep{asplund09}. The error bars account for 1-$\sigma$ errors. The $[{\rm Ne}/{\rm Fe}]$ error bar of LP\,40$-$365 includes the systematic uncertainty, as it anti-correlates with respect to the other elements. \citet{kepleretal16} only determined an upper limit for the Fe abundance of their oxygen white dwarf, thus, the $[{\rm Z}/{\rm Fe}]$ values should be interpreted as lower limits. \label{f:fig05}}
\end{figure}

In Fig.\,\ref{f:fig05}, we compare the abundance pattern\footnote{We define $[{\rm Z}/{\rm Fe}] \equiv \log{({\rm Z}/{\rm Fe})} - \log{({\rm Z}/{\rm Fe})}_{\odot}$ as is conventional. } of LP\,40$-$365 to other white dwarfs exhibiting photospheric metals. As noted by \citet{vennes17}, the composition of LP\,40$-$365 is strikingly different from metal-polluted white dwarfs that  are contaminated by disrupted rocky planetesimals \citep{gaensicke12}, as well as from other low-mass white dwarfs exhibiting photospheric trace metals \citep{hermesetal14-1, gianninasetal14-1}. The O-rich white dwarf identified by \protect\citet{kepleretal16} bears some resemblance to LP\,40$-$365 (i.e.\ large O, Ne, and Mg abundances), however, only four elements have so far been detected in that star (O, Ne, Mg, Si). 

We also confirm LP\,40$-$365 is extremely rich in $\alpha$ elements with respect to the Sun \citep{asplund09}, with  $[\alpha/{\rm Fe}] \equiv [({\rm O}+{\rm Mg})/{\rm Fe}] = 4.6 \pm 0.1$. In our best-fit model, Ne alone accounts for $\simeq 1/3$ of the atmosphere by mass. O, Mg, and the other trace elements make up for 2\%, and 0.4\%, and $\simeq 0.1$\% of the mass, respectively. Notably, Na and Al are enhanced with respect to Solar ratios by factors of $\simeq 100$ and 40, respectively, and also the four iron-group elements we identify (Ti, Cr, Mn, and Ni) are between 2 and 8 times more abundant then Fe with respect to their Solar values.

\subsection{Radial velocity and rotation speed} \label{sec:three-2}
\citet{vennes17} established LP\,40$-$365 as a single star. Our spectroscopic and photometric observations also show no radial velocity variation and a flat light curve, in agreement with their analysis. Cross-correlating our best-fit model with the WHT/ISIS spectra acquired on 2017 July 6, we measure the radial velocity shifts with an accuracy of 2--3\,km/s (listed in Table\,\ref{t:t04}). We note a systematic difference between the radial velocity measurements obtained from the blue and red arms separately, which is caused by a poorer wavelength calibration in the blue due to the small number of strong lines in the CuNe+CuAr arcs. By averaging these measurements, we estimate a radial velocity shift of $v_{r} = 499 \pm 6$~km\,s$^{-1}$, where the associated uncertainty accounts also for the systematic offset in the calibration. From the combined spectra, we estimated a rotation velocity of $v_{\rm rot} \sin{i} < 50$~km\,s$^{-1}$.  Both results are in agreement with those of \citet{vennes17}. 
\begin{table}
\caption{Radial velocity measurements from the WHT/ISIS spectra taken in July 2018. \label{t:t04}}
\centering
\begin{tabular}{cccc}
\hline
\hline
\multicolumn{2}{c}{Blue Arm (4500--5400\,\AA)}& \multicolumn{2}{c}{Red Arm (8100--9000\,\AA)}\\
${\rm MJD} - 57940$  & $v_{\rm r}$ & ${\rm MJD} - 57940$ & $v_{\rm r}$ \\
(days) & (km/s) & (days) &(km/s) \\
\hline
1.1312841& $499.27 \pm 1.72$  & 1.1317123&$498.90 \pm 2.30$\\
1.1384947& $500.73 \pm 1.60$  & 1.1389230&$489.82 \pm 2.32$\\
1.1457054& $500.74 \pm 1.82$  & 1.1461220&$488.83 \pm 2.95$\\
1.1619670& $512.97 \pm 2.06$  & 1.1619554&$493.26 \pm 2.49$\\
1.1691776& $505.86 \pm 1.97$  & 1.1691660&$490.85 \pm 2.68$\\
1.1763883& $506.11 \pm 1.83$  & 1.1763651&$491.12 \pm 2.23$\\
1.1868281& $504.96 \pm 1.67$  & 1.1868396&$492.04 \pm 2.38$\\
1.1940387& $506.29 \pm 1.91$  & 1.1940387&$498.11 \pm 2.14$\\
1.2012609& $501.22 \pm 1.66$  & 1.2012494&$494.96 \pm 2.70$\\
\hline
\end{tabular}
\end{table}
\section{The progenitor of LP\,40$-$365} \label{sec:four}
\begin{figure}
\centering
\includegraphics[width=\linewidth]{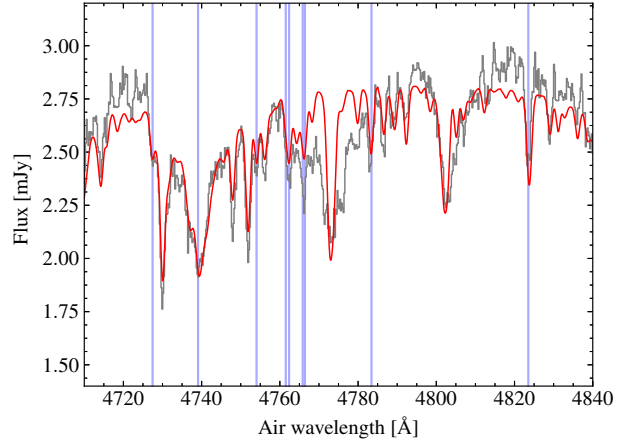}
\caption{The 1-\AA\ resolution spectrum in the 4720--4860\,\AA\ region containing the strongest Mn lines. The best-fit model (red) is overlaid onto the observed spectrum (gray). The vertical lines display the wavelengths of Mn lines. The observed spectrum is blue-shifted into the rest frame.\label{f:fig06}}
\end{figure}
The exotic nature of LP\,40$-$365 has raised the question of whether any established supernova model can explain both its peculiar composition and the high space velocity \citep{vennes17}. \citet{vennes17} speculated these characteristics to be a compelling argument in favor of the ejection of LP\,40$-$365 from a single-degenerate binary system, in which a subluminous thermonuclear explosion failed to entirely disrupt the white dwarf, as outlined in the hydrodynamical simulations of \citet{jordan12} and \citet{kromer13}. In these simulations, the supernova partially disrupts the white dwarf core due to significant expansion of the star during deflagration, thus an ``unburnt'' remnant is formed. Such supernova remnants are expected to be rich in $\alpha$ and iron-group elements, due to the fall-back of nuclear ashes, and predicted to span across a wide range of masses, depending on the internal structure of the progenitor and the explosion mechanism \citep[][]{fink14,kromer15}. 

A key result of our spectroscopic analysis is the detection of photospheric Mn (Fig.\,\ref{f:fig06}). We measure $[{\rm Mn}/{\rm Fe}] = 0.82 \pm 0.18$, which we interpret as the clear link between LP\,40$-$365 and single-degenerate SNe\,Ia. Intense nucleosynthesis of this element is only expected to take place in high density environments \citep[$\rho \gtrsim 2 \times 10^{8}$\,g\,cm$^{-3}$;][]{seitenzahl13a} through the production and subsequent decay of $^{55}$Co. \citet{seitenzahl13a} have studied the Mn production from core-collapse supernovae and SNe~Ia. Their results show that Mn is mostly produced in SNe~Ia and, furthermore, the Mn-to-Fe ratio is expected to be super-Solar ($[{\rm Mn}/{\rm Fe}] > 0$) in near-Chandrasekhar mass explosions ($M \gtrsim 1.22$\,M$_{\odot}$) and sub-Solar ($[{\rm Mn}/{\rm Fe}] < 0$) in sub-Chandrasekhar mass explosions.  As it is commonly assumed that single-degenerate SN\,Ia have near-Chandrasekhar mass white dwarfs \citep{seitenzahl13a}, the large [Mn/Fe] we measure is  compatible with LP\,40$-$365 having experienced extreme conditions, leading to the nucleosynthesis of this element in a high density environment. Thus, this evidence strongly suggests a connection with a single-degenerate SN\,Ia.

\subsection{LP\,40$-$365 in the context of SN\,Iax} \label{sec:four-1}

\begin{figure}
\centering
\includegraphics[width=\linewidth]{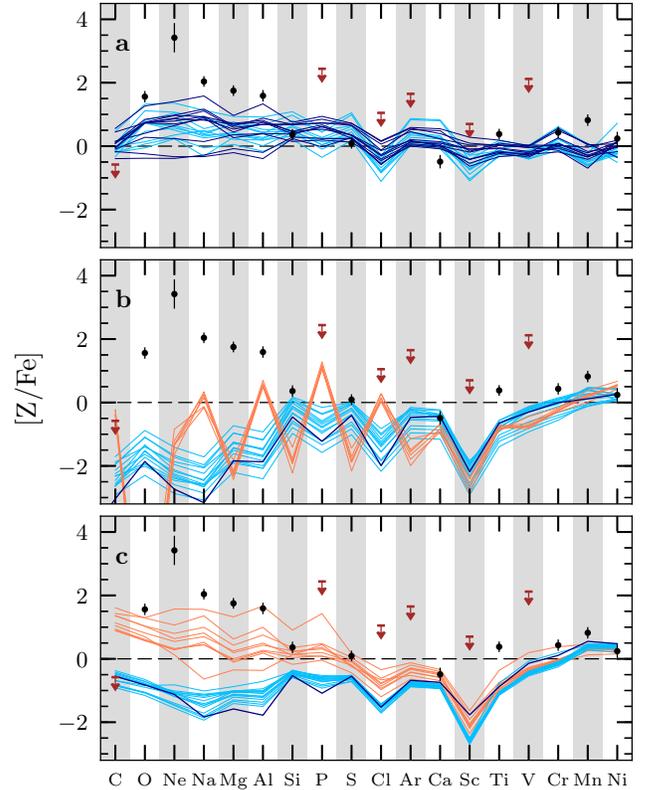}
\caption{The atmospheric composition of LP\,40$-$365 is displayed for detected elements (black circles with error bars) and upper limits (downward pointing brown arrows), which we compare to to theoretical nuclear yields for (a) core-collapse supernovae, (b) SNe~Ia, and (c) SNe~Iax. (a): core-collapse supernova yields by \citet{nomoto06} for metal poor ($Z = 0.004$; light blue curves) and Solar metallicity (dark blue) progenitors. (b): SN~Ia yields from 3-D pure deflagration computations \citep[coral curves;][]{roepke06}, 3-D delayed detonations \citep[light blue curves;][]{seitenzahl13b} and 2-D delayed detonation models \citep[dark blue curve;][]{townsley16}. (c): SNa~Iax yields from 3-D delayed detonation C/O white dwarf \citep[light blue curves;][]{fink14} and C/O/Ne white dwarf models \citep[dark blue curves;][]{kromer15}, and bulk composition of unburnt remnants \citep[coral curves;][]{fink14}.\label{f:fig07}}
\end{figure}

Given that our spectral analysis points towards a He-dominated atmosphere, we investigated the possibility that LP\,40$-$365 was a He (burning) star either in a binary system where the more massive companion underwent a core-collapse supernova or the donor star in a SN~Ia \citep{wang09c}. We compared LP\,40$-$365 to theoretical yields of SNe~Ia as well as core-collapse supernovae (Fig.\,\ref{f:fig07}a-b).  The overall abundance pattern is not well reproduced by the accretion of ejecta from either types of supernova yields. A core-collapse supernova could lead to $\alpha$ enhancement of the companion's atmosphere, however it is a rare occurrence that has only been observed in handful of black-hole X-ray binaries \citep[][]{israelian99}. The existence of a small star such as LP\,40$-$365 in a core-collapse supernova progenitor appears unlikely \citep[SN\,Ib/Ic;][]{zapartas17}. Furthermore, the [Mn/Fe] measurement would contrast with the core-collapse supernova scenario \citep{seitenzahl13a}.

Hydrodynamical simulations of the interaction between core-collapse supernovae and SNe~Ia with their companion's atmosphere show that mostly iron-group elements with low kinetic energies would be deposited onto the companion \citep[Cr, Mn, Fe, Co, and Ni;][]{marietta00,pan12,liu13,liu15}. 
If LP\,40$-$365 had accreted just the iron-group elements by capturing supernova ejecta, an alternative route to $\alpha$-enhancement would be self-pollution through the CNO cycle. Such phenomenon is observed in evolved stars like helium-rich subdwarfs \citep[][]{heber16} or in extreme helium stars \citep[e.g.][]{kupfer17,jeffery17}, however, these stars are typically hotter than LP\,40$-$365 ($\simeq 20\,000$--$40\,000$~K) and do not display Ne abundances as large as that of LP\,40$-$365. We note also that composition,  $T_{\rm eff}$, and $\log{g}$ do not appear to match the properties and evolution of a hypothetical He star that was ejected from a binary system as the consequence of a SN~Ia explosion \citep[][]{liu13,pan13}.

In their interpretation of the $\alpha$-enhancement with respect to iron-group elements, \citet{vennes17} stressed that theoretical predictions for the bulk composition of unburnt supernova remnants may not be representative of the atmospheric composition, suggesting that gravitational settling of heavier elements would take place after the explosion. This interpretation is likely true, but we suggest that the effect of gravitational settling, initially very fast, may have at this stage slowed down \citep[diffusion of metals in cool He-atmospheres is of the order of a hundred million years;][comparable to the travel time of LP\,40$-$365 from below the Galactic plane; \citealt{vennes17}]{paquette86} due to the contribution of other factors like: convective mixing, radiative levitation, and rotational support \citep[e.g.][]{istrate16}. These phenomena are suggested to take place in other low-mass white dwarfs, explaining the metal traces in their atmospheres \citep{hermesetal14-1}. We note that the abundance pattern of LP\,40$-$365 appears to be broadly consistent with -- although not quite exactly the  same as -- the bulk composition predicted for unburnt remnants produced in pure-deflagration simulations (Fig.\,\ref{f:fig07}c), which could be compatible with the dredge-up of nuclearly-processed core material. From our comparison, we found the closest resemblance with the bulk composition of supernova remnants experiencing the most vigorous explosions, which only leave 0.1--0.2\,M$_{\odot}$ remnants. However, we stress that theoretical predictions disagree with three key features of our spectral analysis, i.e.\ the large Ne abundance and, moreover, the He-dominated atmosphere as well as the absence of C. \citet{vennes17} speculated that a C-poor remnant like LP\,40$-$365 could form via the explosion of a hybrid C/O/Ne white dwarf \citep{denissenkov13,bravo16}. Instead, we note that CO-rich material needs to be accreted in order to ignite such a white dwarf \citep{denissenkov15,kromer15,willcox16}; thus, the unburnt remnant could still contain large amounts of C below the photosphere.

While the existence of hybrid white dwarfs themselves and their fate as SNe~Ia are still debated \citep{brooks17}, promising hydrodynamic simulations have shown that more-massive ($M \gtrsim 1.1$\,M$_{\odot}$), O/Ne white dwarfs could also explode as faint SNe~Ia without collapsing into a neutron star, but leaving an unburnt remnant \citep{jones16}. Although the visible display of such thermonuclear explosion is yet unknown, the surviving remnant is expected to have low-C abundances and be mainly made of O, Ne, and iron-group elements.

\section{Summary and conclusions} \label{sec:seven}

We have presented new photometry and spectroscopy of the recently identified hyper-runaway white dwarf,  LP\,40$-$365 (aka GD\,492). This unique star has been proposed as the evidence for a failed SN~Ia explosion in our Galaxy \citep[][]{vennes17}.

From our spectroscopic analysis, we find that a He-dominated atmosphere is also consistent with the data, but it is in contrast with the O/Ne atmosphere proposed by \citet{vennes17}. Our best-fit model reproduces most  of the observed features in the data over the 3200--9100\,\AA\ range, which extends previous observations to bluer and redder wavelengths. We find that LP\,40$-$365 is cool, $T_{\rm eff} \simeq 8900$~K, and has a low surface gravity, $\log{g} \simeq 5.5$, implying its atmosphere is likely convective. We also confirm that the atmosphere of LP\,40$-$365 is very rich in metals (13 detected atomic species), with a large Ne content ($\sim 33$\% by mass). The next most abundant elements are O and Mg ($\simeq 2$\% and 0.4\,\% by mass, respectively). Owing to the systematic uncertainties in assessing [Ne/He], the Ne content could range between 20--50\% of the atmosphere by mass. Improved observations, with a higher signal-to-noise ratio could help to improve the detection of a possibly weak He\,{\sc i}\,5877\,\AA\ line. Although our synthetic spectrum well reproduces most of the remaining spectral lines as well as the global features of LP\,40$-$365 over the 3200--9100\,\AA\ range (e.g.\ the Mg ionization-edge and the general spectral slope), there are some broad absorption features that have a so far unclear origin. 

Supported by the Mn detection, with $[{\rm Mn/Fe}] = 0.82 \pm 0.18$, we speculate that LP\,40$-$365 formed from a single-degenerate SN~Ia progenitor \citep{seitenzahl13a}. The abundance pattern and atmospheric properties of LP\,40$-$365 appear incompatible with the hypothesis of it being the former donor star in a SN~Ia. However, the scenario proposed by \citet{vennes17}, where LP\,40$-$365 is the unburnt remnant of a failed SN~Ia \citep{jordan12,kromer13}, also contrasts with a He- and Ne-rich, C-poor atmosphere. Alternative scenarios, involving the failed explosion of O/Ne white dwarfs \citep{jones16} await future theoretical investigation.

We conclude that the interpretation of LP\,40$-$365 is still subject to some uncertainties in the spectral analysis, which will be eventually solved in future follow-up observations. Securing high quality observations over the entire spectrum, e.g.\ covering from the near-ultraviolet to the infrared, will provide much stronger constraints on the main atmospheric components and atomic abundances, $T_{\rm eff}$, and $\log{g}$. Its distance, which is currently unknown, will be accurately measured by the ESA {\em Gaia} mission \citep[][]{gaia16}, which will also provide a much more precise value on the proper motion. These astrometric data, in combination with binary population synthesis, will especially help to better constrain the nature of the progenitor of LP\,40$-$365 and its evolutionary history.

We stress the importance of identifying and characterizing objects like LP\,40$-$365 as they provide crucial observational constraints for the theoretical framework of thermonuclear supernova explosions.

\acknowledgments

Based on observations collected at Copernico telescope (Asiago, Italy) of the INAF - Osservatorio Astronomico di Padova. The William Hershel Telescope and its service programme are operated on the island of La Palma by the Isaac Newton Group of Telescopes in the Spanish Observatorio del Roque de los Muchachos of the Instituto de Astrofísica de Canarias (prog. ID: W/2017A/25, W/2017A/30, and SW2017a12). The Liverpool Telescope is operated on the island of La Palma by Liverpool John Moores University in the Spanish Observatorio del Roque de los Muchachos of the Instituto deAstrofisica de Canarias with financial support from the UK Science and Technology Facilities Council. 

The research leading to these results has received funding from the
European Research Council under the European Union's Seventh Framework
Programme (FP/2007-2013) / ERC Grant Agreement n. 320964 (WDTracer).

Support for this work was provided by
NASA through Hubble Fellowship grant \#HST-HF2-51357.001-A,
awarded by the Space Telescope Science Institute, which is 
operated by the Association of Universities for Research in Astronomy,
Incorporated, under NASA contract NAS5-26555.

\facilities{1.82-m Copernico(AFOSC), 4.2-m WHT(ISIS), 2-m LT(RISE)}

\end{document}